\documentclass[aps,floatfix,twocolumn,superscriptaddress]{revtex4}
\usepackage{amsmath}
\usepackage{epsfig}
\usepackage{dcolumn}
\usepackage{verbatim}
\usepackage{hyperref}
\usepackage{txfonts}
\usepackage{times}
\usepackage{mathptmx}

\newcommand{\gtsim}{\protect\raisebox{-0.5ex}{$\:\stackrel{\textstyle >}
        {\sim}\:$}}

\newcommand{\kT}{k_{\mathrm{B}}T}

\newcommand{\hint}{h^{\mathrm{(int)}}} 
\newcommand{\hext}{h^{\mathrm{(ext)}}} 

\DeclareMathOperator{\rnd}{rnd}

\begin{document}

\title{Faster strain fluctuation methods through partial volume updates}
\author{Sander Pronk}
\affiliation{Department of Bioengineering,
    University of California, Berkeley,
    94720 Berkeley, CA,
    USA
}

\author{Phillip L. Geissler}
\affiliation{Department of Chemistry,
    University of California, Berkeley,
    94720 Berkeley, CA,
    USA
}

\begin{abstract}
Elastic systems that are spatially heterogeneous in their mechanical response pose special challenges for molecular simulations.  Standard methods for sampling thermal fluctuations of a system's size and shape proceed through a series of homogeneous deformations, whose magnitudes can be severely restricted by its stiffest parts.  Here we present a Monte Carlo algorithm designed to circumvent this difficulty, which can be prohibitive in many systems of modern interest. By deforming randomly selected subvolumes alone, it naturally distributes the amplitude of spontaneous elastic fluctuations according to intrinsic heterogeneity. We describe in detail implementations of such ``slice moves'' that are consistent with detailed balance. Their practical application is illustrated for crystals of 2D hard disks and random networks of cross-linked polymers.
\end{abstract}

\maketitle

\section{Introduction}

Modern intersections of chemistry, biology, and materials science focus attention on systems that are substantially nonuniform in their spatial organization; examples include interfaces, structural elements of the cell such as the cytoskeleton, and systems undergoing phase transitions. Tools of statistical mechanics that could help clarify their structure and function often do not apply straightforwardly or efficiently in the face of such heterogeneity. This paper concerns a class of computational methods that suffer in this way.

Specifically, we address methods for simulating shape fluctuations of elastic materials. Pioneered by Parrinello and Rahman\cite{Parrinello1980} in the context of molecular dynamics, these approaches extended constant pressure simulation techniques and\cite{Ray1984,Ray1985}, like their predecessors, opened the door for novel computational studies of phase transitions\cite{Najafabadi1983,Lill1994,Wojciechowski1988,Branka1993}. The basic idea of these approaches is simple to understand: treat the parameters of a system's overall geometry as fluctuating dynamical variables, on the same footing as molecular coordinates. In practice, it is convenient to isolate changes in box size and shape by introducing scaled (reduced) coordinates, $\bar{r}_i = h^{-1}_{ij} r_j$ (using Einstein summation convention), where $r^i_j$ is the $j$-coordinate of the position vector of atom $i$ and $h_{ij}$ is a matrix of $d$ lattice vectors defining the periodically replicated $d$-dimensional box geometry. Parrinello and Rahman constructed a Lagrangian with fictitious terms involving $h_{ij}$ and its time derivatives, allowing dynamical simulations of a system with fluctuating shape. One can similarly use Monte Carlo simulations to sample the components of $h_{ij}$ from a Boltzmann distribution\cite{Ray1984}.

The problem with applying these methods to heterogeneous materials is also simple to understand. When $h_{ij}$ changes, so do the physical positions of all atomic coordinates. For example, if one of the basis vectors in $h_{ij}$ defining a rectangular simulation cell is scaled by some factor, the corresponding components of all position vectors $r_i$ become multiplied by the same factor. The stiffness of the resulting motion is determined by the resistance of molecular interactions to these scale and shear transformations. Sampling efficiency is thus determined by the proverbial ``weakest link'': If even a small part of the system strongly resists deformation, then a simulation must await rare, transiently softening fluctuations in local structure that facilitate changes in overall geometry. A system featuring many locally stiff regions becomes nearly intractable, since the likelihood of many rare local fluctuations occurring simultaneously is extremely small.

Our attention to these methodological issues is driven by an interest in the polymer networks that determine elastic properties of living cells\cite{Gardel2004,Wilhelm2003}. As a crude but illustrative model of these cytoskeletal materials, consider a collection of semi-flexible filaments, placed and oriented at random on a two-dimensional plane, that are permanently cross-linked wherever they intersect\cite{Head2003a}.  In this case spatial variations in cross-link density effect substantial variations in local stiffness. Even modest global strains are typically not tolerated by the densest regions of the network. A Monte Carlo simulation of such a model could achieve a reasonable acceptance rate only by using very small displacements in system geometry. As a result, relaxation would proceed quite sluggishly.

In this paper we present a technique that can remove these difficulties by allowing heterogeneous deformations. By transforming only part of a system, we avoid hinging fluctuations of the system as a whole on its stiffest parts. We consider such motions as trial moves in a Metropolis Monte Carlo scheme. We term these as ``slice moves'', since they proceed by choosing slices of a system that deform, leaving the remainder of the system internally unaffected. In section \ref{slicesect}, we introduce the method in detail for both constant pressure ($NPT$) simulations and constant stress simulations, paying careful attention to the requirement of detailed balance. We illustrate the method in section~\ref{simulationsect} through application to the elasticity of crystals of 2-dimensional hard disks, and random networks of cross-linked semi-flexible polymers, and in section~\ref{conclusionsect} we conclude.

\section{Fast Sampling through Partial Volume Moves}
\label{slicesect}

\begin{figure}
\centering
\includegraphics[width=0.95\linewidth]{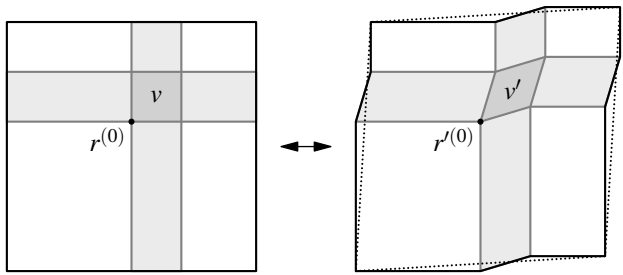}
\caption{A partial volume move. The area $v$ transforms to $v'$, co-transforming the shaded areas. The rest of the simulation box remains unchanged; the new periodically replicating simulation box boundaries are shown as dotted lines.}
\label{slice}
\end{figure}

The basic flaw of conventional strain sampling techniques, when applied to nonuniform systems, is their global nature.  We localize strain moves in Monte Carlo simulations by choosing thin slices of a system, outside of which intermolecular geometries are undisturbed. Fig.~\ref{slice} illustrates such a partial volume move. In this two-dimensional example, subvolumes to be deformed are defined by two intersecting swaths. As a trial move, we deform the region $v$ shared by both slices, producing a new subvolume geometry $v'$. The requirement that regions outside the two slices remain undeformed then uniquely determines transformations within the remaining slice regions (i.e., within one but not both swaths). By choosing the slices' locations and widths at random, we can in effect sample around problematically rigid parts of a configuration.  

\begin{figure}
\centering
\includegraphics[width=0.85\linewidth]{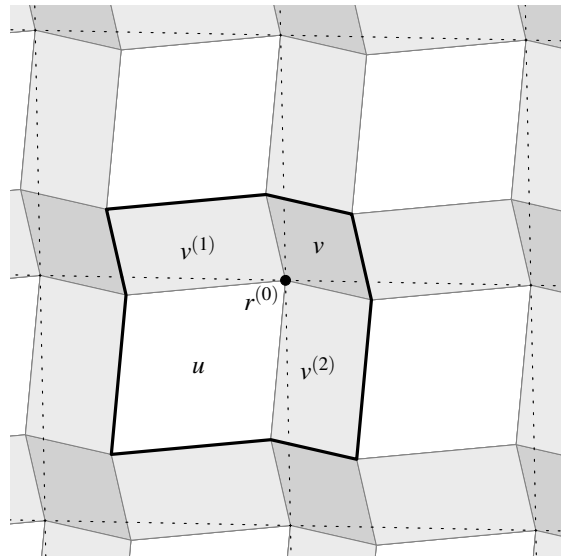}
\caption{The connecting subvolumes $v^{(1)}$ and $v^{(2)}$, and the coordinate shifts and periodic boundary conditions during a slice move. The dashed lines show the original periodic boundary box outlined by $h_{ij}$ (already shifted to $r^{(0)}$ at its origin), while the base periodic image during the slice move is shown with the thick lines. }
\label{periodic}
\end{figure}

Algorithmically, such a ``slice move'' proceeds as follows:
\begin{enumerate}

\item Select a particle at random, whose position $r^{(0)}$ serves as an anchor for the primary deformation subvolume $v$.

\item Select random parallelepipeds $v$ and $v'$ defining initial and final geometries of the subvolume.

\item Determine additional parallelepipeds $v^{(1)}$, $v^{(2)}$, etc., that connect $v$ with its periodic images (See Fig.~\ref{periodic}). These regions, together with $v$, form the intersecting slices that will be deformed.  Repeat for $v'$.

\item Calculate each particle's position in the deformed trial state according to strains applied to the region in which it resides.

\item Evaluate the change in internal energy $\Delta U$ (accounting for the change in periodic boundary conditions) and the work $W$ associated with external forces.

\item Accept or reject the trial move with a probability determined by the 
total change in energy relative to $\kT$

\end{enumerate}

We will describe two variants of such a trial move. The simpler version involves only the limited class of transformations that switch between rectangular system geometries, for which the shape matrices describing $v$, $v'$, $h$, etc. are all diagonal. The more general, and in practice much more complicated version, includes the possibility of shear deformations as well.

Different periodic images of a particle may move differently in the course of a partial volume move.  Detailing the algorithm is therefore greatly simplified by a careful and specific choice of images. Fig.~\ref{periodic} illustrates how we select among each particle's set of periodically replicated coordinates, according to the subvolume it occupies.

First, we require that each subvolume ($v$, $v^{(1)}$, $v^{(2)}$, and the unperturbed region $u$) is not fragmented across system boundaries. Since the subvolumes are themselves repeated in space, this criterion does not by itself uniquely specify a choice of particle images. We further choose that the un-fragmented regions are adjacent in a particular way: the subvolumes $v^{(1)}$, $v^{(2)}$, and $u$ must all contact the anchor point $r^{(0)}$. Note that this scheme results in a collection of particle coordinates that do not lie within the boundary of a single simulation cell. Finally, we translate all particles uniformly so that the anchor point $r^{(0)}$ lies at the origin. While geometrically straightforward, this set of operations carries a nontrivial computational overhead. We describe an efficient implementation in Appendix~\ref{detslice}.

\subsection{Scaling slice moves in a rectangular simulation box}
\label{rectslicesect}

\begin{figure}
\centering
\includegraphics[width=0.95\linewidth]{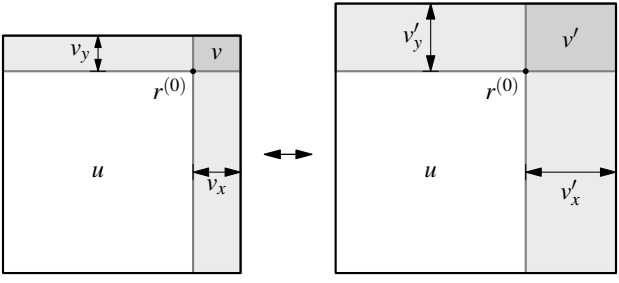}
\caption{A partial scaling move in a rectangular box. For notational simplicity we label co-transforming regions $v_j$ along directions $j$ as $v_x$ (for $j=1$) and $v_y$ (for $j=2$). Corresponding regions in the deformed system are similarly labeled $v_x'$ and $v_y'$ rather than $v_1'$ and $v_2'$, respectively.}
\label{bulkrect}
\end{figure}

The simplest slice move modifies only the scale of rectangular slices along the corresponding lattice vectors, as shown in Fig.~\ref{bulkrect}.  Such a move effects a change in system volume and aspect ratio, but does not change the relative directions of lattice vectors defining periodic boundary conditions.  Here, we will take the box matrix $h_{ij}$ to be purely diagonal, both before and after the distortion. \footnote{For $NPT$ simulations we are not restricted to choosing a rectangular box, but can choose any unit cell that is scaled by a vector $l_i$; the rectangular case was chosen here for notational convenience.}

The width of initial and final slices, together with the slice origin, completely specify an instance of this partial volume move.  We define $v_i$ as the length of subvolume $v$ in direction $i$, and $v_i'$ as its length in the trial configuration. The deformation $s_i$ is simply determined by the ratio of these widths,
\begin{equation}
v'_i = s_{i} v_i.
\label{simplescaling}
\end{equation}
Recall that the slice origin $r^{(0)}_i$ is assigned to be the location of a randomly selected particle. By construction, the lengths $u_i$ of the undeformed region $u$ do not change during a slice move. The box matrix for the trial configuration is therefore given by
\begin{equation}
h'_{ij} = 
\left( u_i + v'_i \right)\delta_{ij}.
\label{newalpharect}
\end{equation}

Accounting for this deformation, the reference frame translation placing $r^{(0)}$ at the origin, and the choice of periodic images depicted in Fig.~\ref{periodic}, we can write the position of particle $i$ in the trial configuration as
\begin{equation}
r'_i = \begin{cases}
    s_i  r_i 
 & 
    \textrm{if }  r_i > 0,
\\[0.3ex]
    \phantom{s_i} 
     r_i  & 
 \textrm{otherwise.}
\end{cases}
\label{newcoordrect}
\end{equation}

In practice, the partial volume transformation is more conveniently performed using reduced coordinates $\bar{r}_i$. In this representation the coordinates of particles in the unperturbed region $u$ change even though their physical arrangements do not:
\begin{equation}
\bar{r}'_i = 
\begin{cases}
\Delta \hint_{ij}  \bar{r}_j
 & 
  \textrm{if }  \bar{r}_j > 0,
\\[1.0ex]
\Delta \hext_{ij}  \bar{r}_j
 & 
 \textrm{otherwise.}
\end{cases}
\end{equation}
Here, the relative deformation matrices $\Delta \hint_{ij}$ and $\Delta \hext_{ij}$ are given by
\begin{subequations}
\begin{align}
\Delta \hint_{ij} &= h'^{-1}_{ik} s_{k} h_{kj},
\\
\Delta \hext_{ij} &= h'^{-1}_{ik} h_{kj}.
\label{boxmatrices}
\end{align}
\end{subequations}

In a Metropolis Monte Carlo simulation, the probability $P_{\mathrm{acc}} ( \Gamma \to \Gamma')$ with which a partial volume move from microstate $\Gamma$ to microstate $\Gamma'$ should be accepted is dictated by the requirement of detailed balance:
\begin{equation}
P_{\mathrm{acc}} ( \Gamma \to \Gamma') = \min\left[1,
\frac{\rho_{\rm equ}(\Gamma')}{\rho_{\rm equ}(\Gamma)}
\frac{P_{\mathrm{gen}} ( \Gamma | \Gamma' ) }
	 {P_{\mathrm{gen}} ( \Gamma' | \Gamma )}
\right]
     \label{detailed_balance}
\end{equation}
Here, $\rho_{\rm equ}(\Gamma)$ is the equilibrium weight of microstate $\Gamma$ in the thermal ensemble of interest; and $P_{\mathrm{gen}} ( \Gamma' | \Gamma )$ is the conditional probability distribution for generated trial configurations $\Gamma'$, given the original configuration $\Gamma$.
This generation probability depends on the way in which slice geometries are chosen. Let $\eta \left( v_m, v'_n, r^{(0)}_o\right)$ be the distribution of parameters specifying a partial volume move.  Since the resulting microstate is uniquely defined by Eqs.~\ref{simplescaling}-\ref{boxmatrices}, $P_{\mathrm{gen}} ( \Gamma' | \Gamma )$ can be written as a product of $\eta \left( v, v', r^{(0)}\right)$ and Dirac delta functions describing the coordinate transformations
\begin{align}
P_\mathrm{gen}(\Gamma' | \Gamma ) & = 
    \prod_{p=1}^{d}
    \prod_{\langle i\rangle_p}
	\delta\left( r'^{(i)}_p - s_p r^{(i)}_p \right)
    \prod_{[j]_p}
	\delta\left( r'^{(j)}_p - r^{(j)}_p \right)
\nonumber\\
	&\quad\times
\eta \left( v, v', r^{(0)}\right).
  \label{probgenrect}
\end{align}
The notation $\prod_{\langle i\rangle_p}$ indicates a product over all $N_p$ particles whose positions are influenced by partial volume scaling in the direction $p$, i.e., particles lying within the slice that runs perpendicular to the $p^{\rm th}$ lattice vector. Similarly, $\prod_{[j]_p}$ denotes a product over the $N-N_p$ particles whose coordinates are unaffected by scaling in the $p$ direction. Note that the anchor point does not change during the transformation -- for accounting purposes, the corresponding particle lies outside the deformed subvolumes.

If the distribution $\eta \left( v, v', r^{(0)}\right)$ is symmetric with respect to exchange of $v$ and $v'$ (i.e., if the original subvolume $v$ and the distorted subvolume $v'$ are selected in the same way), the ratio of generation probabilities appearing in Eq.~\ref{detailed_balance} evaluates simply to 
\begin{align}
\frac{P_\mathrm{gen}(\Gamma | \Gamma' ) }{P_\mathrm{gen}(\Gamma' | \Gamma ) }
&=
\prod_{p=1}^d s_p^{N_p}
\label{pgenratrect}
\end{align}
For a system held at fixed temperature $T = (k_{\rm B}\beta)^{-1}$ and isotropic pressure $P$, equilibrium probabilities depend on internal energy $E$ as well as the total volume $V$, $\rho_{\rm equ}(\Gamma) \propto \exp[-\beta(E+PV)]$. The corresponding Metropolis acceptance probability for a partial volume move is then
\begin{align}
P_{\mathrm{acc}} ( \Gamma \to \Gamma') & = \min\Bigg[1,
\left( \prod_{p=1}^d 
		\left(\frac{v'_p}{v_p}\right)^{N_p}
		\right)
\nonumber\\
&
  \quad \times
	\exp \left( -\beta \left[ \Delta E  
		    + p \left( \left| h'_{qr} \right| - 
			\left| h_{qr} \right|	
			\right) \right] \right)
\Bigg],
\label{metropolis_pacc_rect}
\end{align}
where $\Delta E$ is the change in internal energy resulting from the trial deformation. $|A_{ij}|$ denotes the determinant of a matrix $A$, so that $|h_{ij}|$ and $|h_{ij}'|$ represent the volumes of original and trial states, respectively. 

A partial volume move closely resembles a conventional global strain move when the deformation subvolume $v$ encompasses the whole system, $v_i = h_{ii}$. In this case all particle coordinates (except those of the slice anchor point) are subjected to scaling in each direction, $N_p=N-1$. The acceptance probability then becomes $\min[1,e^{-\beta \Delta U}]$, with an effective potential $U= E + P V + (N-1)k_{\rm B}T \ln V$, much as in a standard isothermal-isobaric Monte Carlo simulation\cite{Frenkel2002}.

\subsection{Slice moves with shear}
\label{arbitraryshape}

\begin{figure}[!tb]
\centering
\includegraphics[width=0.95\linewidth]{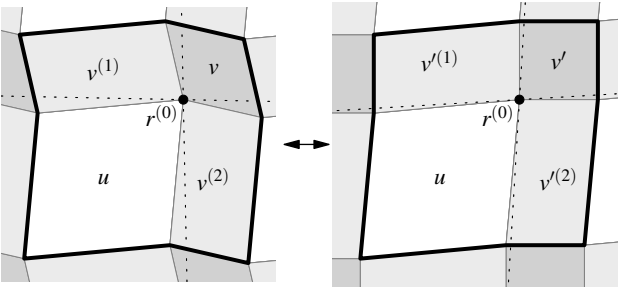}
\caption{A partial volume move with shear components. The thick lines represent the box boundary as used in the slice move.}
\label{shearslice}
\end{figure}

Because lattice vector orientations are invariant under the deformations described in the preceding section, those trial moves do not suffice for simulating shear fluctuations. In this section we present a generalization of slice moves suitable for that purpose. It is tempting to proceed by selecting rectangular slices, as before, and then distorting them into parallelotope shapes (like the deformation sketched in Fig.~\ref{slice}). If restricted to rectangular slices of the initial state, however, a move of this sort is irreversible and therefore inconsistent with detailed balance. Incorporating shear correctly requires the possibility that slices of the initial state also be shaped as parallelotopes (parallelograms in $d=2$ and parallelepipeds in $d=3$), as shown in Fig.~\ref{shearslice}.

The product of a slice move including shear components is a box matrix whose component vectors differ in direction from those of the initial state.  We can therefore no longer treat distortions in different directions as independent deformations. As a mathematical consequence, we require matrices (rather than vectors as in the previous section) to describe subvolume shapes . 

Let $v_{ij}$ be a $d\times d$ matrix whose rows are vectors spanning the edges of the deformation subvolume $v$.  Similarly, the rows of $u_{ij}$ span the edges of the undisturbed region $u$. As sketched in Fig.~\ref{shearslice} periodic boundary conditions demand that
\begin{equation}
h_{ij} = u_{ij} + v_{ij}.
\label{ao1shear}
\end{equation}
Particles residing neither in $v$ nor in $u$ belong to one of several co-transforming subvolumes, whose shape matrices combine one or more rows of $v_{ij}$ with one or more rows of $u_{ij}$.  For the case $d=2$ we denote the two co-transforming regions $v^{(1)}$ and $v^{(2)}$, as shown in Fig.~\ref{shearslice}. (We will discuss the three-dimensional case later.) Subvolume $v^{(1)}$ connects the right edge of $v$ with the left edge of its horizontally replicated periodic image; $v^{(2)}$ connects top and bottom edges of vertically replicated periodic images.  Matrices $v^{(k)}_{ij}$ describing these regions share one row with $v_{ij}$ and one row with $u_{ij}$, \begin{equation}
    v^{(k)}_{ij} = \left\{ 
    \begin{array}{ll}
	v_{ij} & \mathrm{if } j \neq k \\[0.4ex]
	u_{ij} & \textrm{otherwise}. 
    \end{array}
    \right.
\end{equation}
We employ similar definitions for the trial configuration, so that
\begin{align}
h'_{ij} &= u_{ij} + v'_{ij} 
\nonumber\\
&= h_{ij} +
    \left(v'_{ij} - v_{ij} \right)
\end{align}
and
\begin{equation}
v'^{(k)}_{ij}
 = \left\{\begin{array}{ll}
    v'_{ij}  & \textrm{if } j \neq k
    \vspace{0.4ex}
    \\
    u_{ij}  & \textrm{otherwise} 
\end{array}
\right.
\end{equation}

With these definitions we can compactly express deformation matrices describing the strain applied to each subvolume:
\begin{subequations}
\begin{align}
    s_{ij} & = v'_{ik}  v^{-1}_{kj}, \\
    s^{(k)}_{ij} & = v'^{(k)}_{ik}  (v^{(k)})^{-1}_{kj}. \label{sxdef}
\end{align}
\end{subequations}
Particle positions in the trial microstate can finally be written
\begin{equation}
r'_i = \begin{cases}
s_{ij}  r_j 
& 
    \textrm{if } r \textrm{ lies in } v 
\\[0.5ex]
s^{(k)}_{ij}  r_j 
& 
    \textrm{if } r \textrm{ lies in } v^{(k)}
\\[0.5ex]
\phantom{s_{ij}} r_i 
& 
    \textrm{if } r \textrm{ lies in } u
\\
\end{cases}
\label{newcoordshear}
\end{equation}

The generation probability for slice moves including shear is similar to that of the simpler deformations described by Eq.~\ref{probgenrect}:
\begin{align}
P_{\mathrm{gen}} (\Gamma' | \Gamma)= \eta(v, v', r^{(0)}) 
\prod_{p=1}^{d} \Bigg[
 \prod_{j \in {\cal S}(v)}
&	\delta\left( r'^{(j)}_p - s_{pq} r^{(j)}_q \right)
\nonumber\\
 \qquad \quad
\times 
 \prod_{j \in {\cal S}(v^{(1)})}
&	\delta\left( r'^{(j)}_p - s^{(1)}_{pq} r^{(j)}_q \right)
\nonumber\\
 \qquad \quad
\times 
 \prod_{j \in {\cal S}(v^{(2)})}
&	\delta\left( r'^{(j)}_p - s^{(2)}_{pq} r^{(j)}_q \right)
\nonumber\\
 \qquad \quad
\times 
 \prod_{j \in {\cal S}(u)}
&	\delta\left( r'^{(j)}_p - r^{(j)}_p \right)
\Bigg],
\label{probshear}
\end{align}
where ${\cal S}(\alpha)$ denotes the set of $N_{\alpha}$ particles that reside in subvolume $\alpha$. If the subvolumes $v$ and $v'$ are selected independently from the same distribution, as we assumed in Eq.~\ref{pgenratrect}, then the ratio of backward and forward probabilities becomes:
\begin{align}
\frac{P_\mathrm{gen}(\Gamma | \Gamma' ) }{P_\mathrm{gen}(\Gamma' | \Gamma ) }
&= 
	\left| s^{(1)}_{ij}  \right|^{N_{v^{(1)}}} 
	\left| s^{(2)}_{ij}  \right|^{N_{v^{(2)}}}
	\left| s_{ij}  \right|^{N_{v}}
\nonumber\\
&= 
	\left| v'^x_{ik} (v^{(1)})^{-1}_{kj} \right|^{N_{v^{(1)}}} 
	\left| v'^y_{ik} (v^{(2)})^{-1}_{kj} \right|^{N_{v^{(2)}}}
	\left| v'_{ik} v^{-1}_{kj} \right|^{N_{v}},
\label{pgenratioshear}
\end{align}
Detailed balance can therefore be satisfied by accepting these slice moves with a probability:
\begin{align}
P_\mathrm{acc}(v_{ij}, v'_{kl}, r^{(0)}_m)  & = 
\min \Bigg[ 1, 
	\left| s^{(1)}_{mn} \right|^{N_{v^{(1)}}} 
	\left| s^{(2)}_{mn} \right|^{N_{v^{(2)}}}
	\left| s_{mn} \right|^{N_v}  \ \ 
\nonumber\\
& \qquad
    \times
	\exp \Big( -\beta \Big[ \Delta E  + W_{\rm ext} 
	  \Big] \Big)
	\Bigg].
\label{accshear}
\end{align}
As in Eq.~\ref{metropolis_pacc_rect}, $\Delta E$ denotes the change in internal energy resulting from the trial move. The mechanical work $W_{\rm ext}$ against external forces may depend on the box matrices $h_{ij}$ and $h_{ij}'$ in a complicated way if applied stresses are anisotropic.  For the simplest case of constant applied isotropic pressure, this energy takes the familiar form of pressure-volume work, $W_{\rm ext}=P (| h'_{ij}| - | h_{ij}|)$.\footnote{Note that the ensemble of box deformations at constant applied isotropic pressure differs from that at fixed thermodynamic tension. Computing elastic constants from strain fluctuations at constant pressure thus requires a careful accounting of contributions to corresponding compliances from external forces \cite{Ray1984,Wojciechowski1988}.}


Slice moves in three dimensions require a larger and slightly more complicated set of co-transforming subvolumes. We denote these six regions $v^{(kl)}$, where $k$ and $l$ take on integer values corresponding to the three cardinal directions, and $v^{(lk)}$ refers to the same region as $v^{(kl)}$. The parallelepiped $v^{(kk)}$ connects a face of the primary deformation subvolume $v$ with the opposing face of its periodic image in direction $k$, much as for the $d=2$ case. In $d=3$ these subvolumes must themselves be connected by co-transforming regions in order to preserve the undisturbed parallelepiped $u$. The region $v^{(kl)}$, for example, connects periodic images of $v^{(kk)}$ in the direction $l$ (or, equivalently, periodic images of $v^{(ll)}$ in the direction $k$).  Shape matrices for these subvolumes are given by
\begin{equation}
    v^{(kl)}_{ij} = \left\{ 
    \begin{array}{ll}
        v_{ij} & \textrm{if } j \neq k  \textrm{ and }  j \neq l \\
        u_{ij} & \textrm{otherwise}, 
    \end{array}
    \right.
    \label{rx3dext}
\end{equation}

Aside from this enlarged set of subvolumes, coordinate transformations and generation probabilities proceed just as for $d=2$. For example, the deformation matrix for region $v^{(kl)}$ is given by
\begin{equation}
s_{ij}^{(kl)} = v_{im}^{(kl)} \left(v^{(kl)} \right)_{mj}^{-1}.
\end{equation}
The acceptance probability dictated by detailed balance is also simply generalized:
\begin{align}
P_\mathrm{acc}(\Gamma \to \Gamma')  & = 
\min \Bigg[ 1, 
  \left( \prod_\alpha
	\left| s^{(\alpha)}_{ij} \right|^{N_{\alpha}}
\right)
\nonumber\\
& \qquad
    \times
	\exp \Big( -\beta \Big[ \Delta E  + W_{\rm ext} 
	  \Big] \Big)
	\Bigg],
\label{accshear3d}
\end{align}
where the product runs over all subvolumes $\alpha$ (including $v$, $u$, and the co-transforming regions $v^{(kl)}$) with corresponding deformation matrices $s^{(\alpha)}_{ij}$.

\subsection{Selecting $v$ and $v'$}
\label{vchoice}

We have shown that detailed balance is straightforward to achieve with slice moves, provided the selection of subvolumes $v$ and $v'$ is symmetric: \begin{equation}
\eta(v, v', r^{(0)})=\eta(v', v, r'^{(0)}).
\label{symmetry}
\end{equation}
Eq.~\ref{symmetry} is most easily satisfied by choosing the corresponding shape matrices independently, and from the same distribution. Consequently, one's choice of deformed geometry $v_{ij}'$ cannot be biased by the system's current shape $h_{ij}$.  This restriction poses a challenge to efficient sampling. It is advantageous to employ a wide range of subvolume shapes in order to accommodate elastic inhomogeneities that are {\em a priori} unknown; at the same time, typical acceptance probabilities can be very low if $v$ and $v'$ differ substantially. Below we describe a procedure for choosing deformation regions that addresses both of these goals, while respecting the necessity of statistical independence.

A natural method for generating random shape matrices would draw elements from a uniform distribution limited to a certain range $\epsilon$. This approach pits the above goals against one another. Small values of $\epsilon$ discourage generating diverse subvolume shapes. Large values of $\epsilon$ permit significant disparity between independent samples. One simple way of circumventing this dilemma is to vary at random the mean values of distributions from which matrix elements are selected. 

Toward this end we define a symmetric reference matrix
\begin{equation}
\hat{v}_{ij} = \left( 
	\begin{matrix}
	    \rnd(\hat{v}_{xx}^\mathrm{min}, \hat{v}_{xx}^\mathrm{max}) & 
	    \rnd(-\hat{v}_{xy}^\mathrm{min},\hat{v}_{xy}^\mathrm{max}) \\[1.0ex]
	    \hat{v}_{12} &
	    \rnd(\hat{v}_{yy}^\mathrm{min},\hat{v}_{yy}^\mathrm{max}) 
	\end{matrix}
	\right),
\end{equation}
whose elements change stochastically over the course of a Monte Carlo simulation.  Here, $\rnd(a,b)$ denotes a random number uniformly distributed between $a$ and $b$. We employ a given realization of $\hat{v}_{ij}$ as a random offset for selecting both $v_{ij}$ and $v'_{ij}$:
\begin{subequations}
\begin{align}
v_{ij} &= \hat{v}_{ij} + \Delta v_{ij}  =
\hat{v}_{ij} + 
\left( 
	\begin{matrix}
	    \rnd(-\epsilon_{xx}, \epsilon_{xx}) & 
	    \rnd(-\epsilon_{xy}, \epsilon_{yx}) \\[1.0ex]
	    \Delta v_{12} &
	    \rnd(-\epsilon_{yy},\epsilon_{yy}) 
	\end{matrix}
	\right),
\\[1.5ex]
v'_{ij} &= \hat{v}_{ij} + \Delta v'_{ij}  =
\hat{v}_{ij} + 
\left( 
	\begin{matrix}
	    \rnd(-\epsilon_{xx}, \epsilon_{xx}) & 
	    \rnd(-\epsilon_{xy}, \epsilon_{yx}) \\[1.0ex]
	    \Delta v'_{12} &
	    \rnd(-\epsilon_{yy},\epsilon_{yy}) 
	\end{matrix}
	\right).
\end{align}
\end{subequations}
By controlling the ranges $\epsilon_{xx}$, $\epsilon_{xy}$ and $\epsilon_{yy}$ of variations about the reference geometry, similarity of $v$ and $v'$ can be assured and a reasonable acceptance probability maintained. Note that matrix symmetry allows only three elements of $\hat{v}_{ij}$ to be chosen independently.

It can be demonstrated that this scheme obeys detailed balance for any set of fixed parameters $v_{ij}^\mathrm{min}$, $v_{ij}^\mathrm{max}$, and $\epsilon_{ij}$, so long as slices do not exceed the overall system size (just as conventional constant pressure simulations require volume increments smaller than the system's total volume). This constraint should not be limiting: if elastic heterogeneity calls for slice moves, they will be useful only if typical slices are smaller than natural correlation lengths for strain fluctuations.

\section{Simulations}
\label{simulationsect}

We have implemented slice moves in computer simulations of two model
systems, both to verify that equilibrium ensembles of spontaneous
box deformations are correctly sampled and to demonstrate improved
efficiency for elastically heterogeneous systems.

\subsection{Validation: Hard disk solids in two dimensions}

The elastic properties of two-dimensional crystals comprising hard disks have been calculated with high precision\cite{Wojciechowski1988,Bates2000,Sengupta2000,Wojciechowski2003} in efforts to assess the possibility of a KTHNY transition\cite{Nelson2002,Chaikin1995}. (A sufficiently low Young's modulus signals instability to the creation of dislocations, implying a two-stage, continuous transition to the disordered fluid phase.) Here we use those results, obtained using conventional approaches, as benchmarks for validating our new methods.

\begin{figure}
\centering
\includegraphics[width=0.85\linewidth]{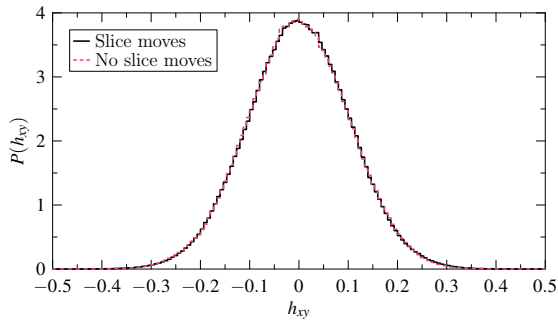}
\caption{Probability distribution function $P(h_{xy})$ for the shear component $h_{xy}$ of the box shape matrix for the two-dimensional hard disk system at pressure $P=10 \  \kT/\sigma^2$.}
\label{hdhxyhist}
\end{figure}

We have simulated systems of $N=780$ hard disks at constant pressure $P$ with periodic boundary conditions. Pairwise interactions forbid interparticle separations smaller than the particle diameter $\sigma$ but otherwise do not bias spatial arrangements. For $P \gtsim 9\ \kT/\sigma^2$ the equilibrium state is a crystalline solid\cite{Alder1962}, whose bulk modulus $B$ and effective shear constant $\mu_{\rm eff}$ we determine from distributions of spontaneous fluctuations in Lagrangian strain\cite{Ray1984,Ray1985,Wojciechowski1988,Sengupta2000}. Fig.~\ref{hdhxyhist} illustrates the breadth of strain fluctuations for the specific case $P= 10\ \kT/\sigma^2$.

We have calculated $B$ and $\mu_{\rm eff}$ for hard disk solids at several pressures. In each case we performed one simulation using slice moves and one using exclusively conventional methods. Our results are shown in Table \ref{hdrestab}, along with previously reported values for systems at similar conditions.  Simulations with and without slice moves agree well, yielding results for most pressures that lie well within error margins. At $P=10$ and $P=11$ there are some differences in the bulk modulus $B$ that might be explained through the moderate softness of the crystal itself; we speculate that the slice moves enable better sampling of local defects that lead to elastic heterogeneities, leading to slightly different bulk moduli\cite{Bates2000}, and unrealistically low error estimates in the case of no slice moves. Aside from that, the results match the most accurate results published elsewhere.

\begin{table}
\begin{tabular}{l r  r@{}l     r@{}l    r@{}l}
 method & $N$ & 
 $P$    & & 
 $B$    & & 
 \multicolumn{2}{c}{$\mu_{\rm eff}$}\\
\hline
\hline
sf, slice moves & 780               & 9&.545    & 42&(1)        & 15&.3(3) \\
sf, no slice moves & 780            & 9&.545    & 42&.4(5)      & 15&.0(2) \\
\hline
sf, slice moves & 780               & 10&.0     & 48&.4(7)      & 21&.96(9) \\
sf, no slice moves & 780            & 10&.0     & 51&.0(2)      & 21&.78(5) \\
sf \cite{Wojciechowski2003} 
                         & 896  & 10&.0     & 49&.2(8)      & 21&.9(3) \\
\hline
sf, slice moves & 780               & 11&.0     & 60&.8(8)      & 26&.3(1) \\
sf, no slice moves & 780            & 11&.0     & 57&.9(3)      & 26&.1(2)  \\
\hline
sf, slice moves & 780               & 11&.6     & 69&(1)        & 28&.89(9) \\
sf, no slice moves & 780            & 11&.6     & 67&(1)        & 28&.9(3) \\
sf \cite{Wojciechowski2003} 
                                & 896  & 11&.6     & 68&(2)        & 28&.8(4) \\
stress-strain \cite{Wojciechowski2003} 
                                & 7020 & 11&.6     & 67&(1)        & 28&.8(3) \\
\hline
sf, slice moves & 780               & 13&.0     & 85&(1)        & 35&.8(2) \\
sf, no slice moves & 780            & 13&.0     & 85&(1)        & 35&.8(3) \\
\hline
sf, slice moves & 780               & 15&.4     & 118&(2)       & 48&.5(3) \\
sf, no slice moves & 780            & 15&.4     & 119&(2)       & 49&.4(3) \\
sf \cite{Wojciechowski2003} 
                         & 896  & 15&.4     & 118&(1)       & 48&.7(3) \\
stress-strain \cite{Wojciechowski2003} 
                         & 7020 & 15&.4     & 118&(1)       & 49&.2(3) \\
\hline
sf, slice moves & 780               & 23&.1     & 255&(4)       & 104&.9(5) \\
sf, no slice moves & 780            & 23&.1     & 260&(5)       & 104&.0(8) \\
sf \cite{Wojciechowski2003}
                         & 896  & 23&.1     & 251&(4)       & 104&(1)   \\
stress-strain \cite{Wojciechowski2003} 
                         & 7020 & 23&.1     & 252&(3)       & 103&.3(9) \\
\end{tabular}
\label{hdrestab}
\caption{ Elastic constants of 2D hard disk crystals, comparing strain fluctuation (`sf') simulations with and without slice moves, and previous work. $N$ is the number of particles in the system, $P$ is the applied isotropic pressure, $B$ is the bulk modulus, and $\mu_{\rm eff}$ is the effective shear modulus\cite{Sengupta2000}.  Units are in particle diameter $\sigma$ and $\kT$; the number in parentheses represents the standard error in the last digit.}
\end{table}

In terms of computational efficiency, hard-disk crystals represent something of a worst-case scenario for slice moves: the interparticle potential requires little numerical effort to evaluate, elastic response is spatially uniform, and the interactions are extremely short-ranged.  These factors render significant the added computational overhead of a slice move relative to a conventional strain move.  Specifically, the run time is approximately 25\% longer for simulations with slice moves compared to those lacking slice moves but comprising the same total number of strain moves.

\subsection{Cytoskeletal networks}

Slice moves may offer considerable computational savings whenever the resistance of a system to strain varies significantly in space. Here we demonstrate their utility for a model elastic gel inspired by the polymeric framework of living cells. This two-dimensional system comprises a collection of semi-flexible filaments connected by cross-links. For the specific model we consider here, cross-links enforce overlap of two filaments at fixed points along their contours but do not constrain the angle at which they intersect.  A thorough examination of this model's elastic response will be presented in a forthcoming paper.

\begin{figure}
\includegraphics[width=0.88\linewidth]{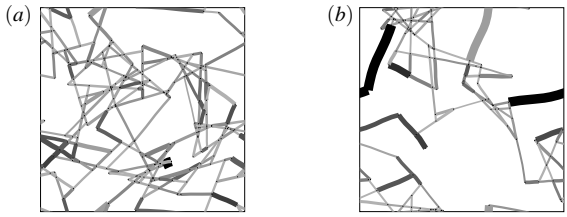}
\caption{Two randomly laid down networks of model actin filaments during a Monte Carlo simulation. The line widths denote instantaneous parallel strain for the segment, while the darkest shades denote higher instantaneous perpendicular strain for the segment. The sizes of the systems are $0.25 l_p \times 0.25 l_p \approx 2.5 \ \mu \mathrm{m} \times 2.5 \ \mu \mathrm{m} $, where $l_p$ is the persistence length of actin. The system in $(a)$ shows a medium density system (with average distance between cross-links $l_c=0.014l_p$), while the system in $(b)$ shows a low density system (with $l_c=0.019l_p$). Both have rigid cross-links, and average filament length $0.11 l_p$ before removal of free  end-points.}
\label{conf}
\end{figure}

We construct a particular realization of the network by laying down straight filaments of fixed length, located and oriented at random, until a desired density is achieved. Wherever filaments intersect, they become permanently cross-linked. We will focus on two such configurations, differing in density. Both are shown in Fig.~\ref{conf}. The contour length of a filament segment between two-cross-links is set such that the initial distance between cross-links minimizes the segment's free energy.

Our simulations focus explicitly on fluctuations in the positions and orientations of cross-links, which primarily characterize the elasticity of this model system. In particular, a microstate $\Gamma$ specifies only the configuration of cross-links, including the directions in which filaments pass through them. Thermal undulations of filament segments consistent with $\Gamma$ are integrated out beforehand
according to the statistical mechanics of a worm-like chain\cite{Marko1995}. The ``energy'' $E$ associated with $\Gamma$ thus in fact represents a free energy that accounts for the corresponding variety of chain configurations. It is a highly nonlinear function of cross-link arrangements, due to the inextensibility of a worm-like chain along its contour. These sharp nonlinearities foster heterogeneous stiffness and impede calculations of elastic response.

\begin{figure}
\includegraphics[width=0.95\linewidth]{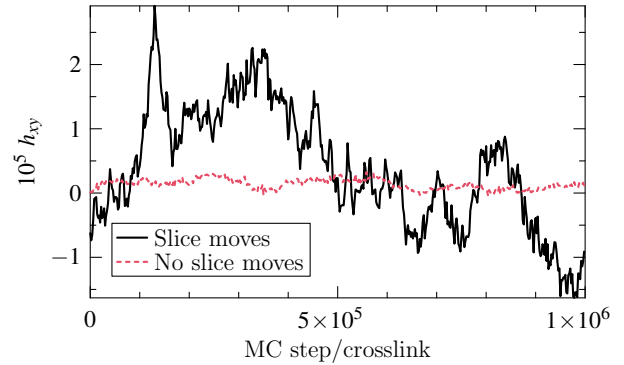}
\caption{Shear component $h_{xy}$ during a MC simulation under zero pressure, with and without slice moves. The simulated network is shown is Fig~\ref{conf}$(a)$. As in Fig.~\ref{bulkrect}, $x$ and $y$ denote directions of the two lattice vectors describing the geometry of this periodically replicated system. 
}
\label{random}
\end{figure}

\begin{figure}
\centering
\includegraphics[width=0.95\linewidth]{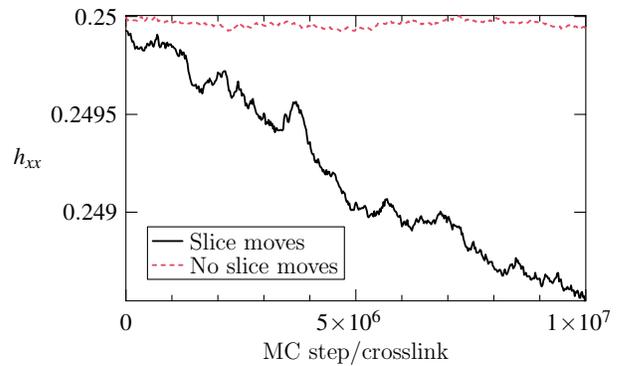}
\caption{Bulk component $h_{xx}$ during a MC simulation under applied pressure for a network undergoing collapse through buckling, with and without slice moves. The simulated network, whose horizontal direction is denoted by $x$, is shown in Fig~\ref{conf}$(b)$.}
\label{dilute}
\end{figure}

Figs.~\ref{random} and \ref{dilute} show results of Monte Carlo simulations for these model networks. They contrast fluctuations and relaxation generated using conventional methods of sampling at constant pressure with those produced by slice moves.  Slice shapes $v_{ij}$ and $v'_{ij}$ were chosen according to the recipe in Section~\ref{vchoice}. By setting $\hat{v}^{\mathrm{min}}_{xx} = 0.05 h_{xx}^{\rm (init)}$, $\hat{v}^{\mathrm{min}}_{yy} = 0.05 h_{yy}^{\rm (init)}$, $\hat{v}^{\mathrm{max}}_{xx} = 0.2 h_{xx}^{\rm (init)}$, and $\hat{v}^{\mathrm{max}}_{yy} = 0.2 h_{yy}^{\rm (init)}$, where $h_{ij}^{\rm (init)}$ is the shape matrix at the beginning of the simulation, we generate deformation subvolumes with dimensions $5\%-20\%$ of the initial system size.  The step size of trial shear deformations, $\epsilon_{xy} \approx 10^{-7}$, was tuned during an equilibration phase of the simulation to establish an acceptance ratio of approximately $0.5$.

Trajectories of spontaneous shear strain fluctuations are plotted in Fig~\ref{random} for the denser network configuration shown in Fig.~\ref{conf}$(a)$. The enhanced efficiency offered by slice moves for sampling thermally accessible strain states is clearly evident.  By itself, the result for conventional, global strain moves provides no warning that it has failed to visit important regions of configuration space. One might therefore be tempted to estimate elastic susceptibilities, which would be orders of magnitude too small, from a severely deficient set of thermal fluctuations. 

The sparser network configuration shown in Fig~\ref{conf}$(b)$ is extremely susceptible to applied pressure. The bulk strain trajectory obtained using slice moves manifests this pliability through a systematic decrease in box size under load. (See Fig.~\ref{dilute}.) With conventional methodology, by contrast, contraction of the network as a whole necessitates deforming its densest regions, at least transiently. Indeed, when restricted to global strain moves, Monte Carlo sampling cannot access compressed states even within $10^7$ sweeps.

Because energy evaluations in network simulations are more numerically taxing than in hard disk simulations, the added overhead for performing slice moves amounts to a scant 1\% increase in run time compared to conventional simulations comprising the same total number of strain moves. This price is clearly outweighed by the dramatic gains in computational efficiency we have demonstrated.  We expect efficiency considerations to similarly favor the use of slice moves for other complex systems that exhibit heterogeneous elasticity. In most physical contexts of interest, evaluating changes in potential energy due to intermolecular interactions will make negligible even the greatest expense brought on by slice moves, namely, determining which subvolume each particle occupies when executing a deformation with non-rectangular slices. Furthermore, rectangular slices should suffice for exploring many types of elasticity, e.g. in systems that are fluid; assigning particles to rectangular subvolumes is numerically inconsequential compared to calculating interaction energies for all but the simplest systems.

\section{Conclusion}
\label{conclusionsect}

We have shown how volume moves in constant-pressure simulations and strain moves in constant-stress simulations can be performed locally, such that intermolecular arrangements in much of a system remain undisturbed. Significant speedup of Monte Carlo simulations is expected for systems that are considerably nonuniform in stiffness. Example simulation results confirm that physically important strain states previously inaccessible as a matter of practice can now be readily explored. 

By facilitating spontaneous strain fluctuations, this methodological advance promises to greatly extend the purview of techniques that assess linear elastic response via the fluctuation-dissipation theorem. Additionally, it provides a new type of collective Monte Carlo move as an alternative to cluster moves\cite{Frenkel2002,Swendsen1987,Whitelam2007}. 

More broadly, it opens doors to applications in the many biophysical and materials contexts that involve spatially varying density (as occurs in a material undergoing a phase transition) and/or composition (as is routine in living cells). 

\section{Acknowledgments}
This work is  supported in part by the California Institute for Quantitative Biosciences, and by the National Science Foundation.

\appendix
\section{Algorithm to determine which slice a point is in}
\label{detslice}

Executing a slice move requires determining the set of particles that reside in each subvolume, before their coordinates can be appropriately transformed (according to Eq.~\ref{newcoordrect} or \ref{newcoordshear}). Performing this task efficiently is straightforward for subvolumes that are rectangular in the reduced coordinate space. For non-rectangular slices, however, it can become both awkward and costly. Here we outline an algorithm that, for most points in a simulation box, reduces the classification problem to checking whether the point lies within a particular rectangle.

The essence of this procedure is to inscribe a rectangle within each subvolume $\alpha$ (where $\alpha \in \{u, v, v^{(x)}, v^{(y)}\}$ in two dimensions). Particle coordinates can be quickly checked against these rectangles. Because useful deformation volumes tend to be small, most particles will fall within the inscribed rectangle of the undisturbed region $u$. Only a small fraction of particles need then be checked against subvolumes' full parallelotope shapes. A systematic procedure for doing so is described below.

Consider a particle located at position $\bar{r}$ in the reduced coordinate system, and a subvolume $\alpha$ centered at position $\bar{c}$ (also in the reduced coordinate system) with shape matrix $\alpha_{ij}$. We first determine which of the particle's periodic images, whose position we denote $\bar{r}^*$, lies nearest $\bar{c}$.  We then compute a new set of reduced coordinates, $\bar{r}'_i = (\alpha^{-1})_{ij}h_{jk}[\bar{r}^*_k-\bar{c}_k]$, referenced to the subvolume shape and translated so that the origin lies at $\bar{c}$.  If $-1/2 \leq \bar{r}'_i \leq 1/2$ for all $i=1,2,\ldots,d$, then the particle resides in $\alpha$.  By ordering subvolumes according to size, and checking particle positions against the largest slices first, we can ensure that most particles are assigned without numerous repetitions of these transformations.


\end{document}